\documentclass[manuscript]{aastex62}
\shorttitle{Probing TTV and its possible origin in TrES-3 system}
\shortauthors{Mannaday et al.}

\usepackage{longtable}
\begin{document}

\title{Probing Transit Timing Variation and its Possible Origin with Twelve New Transits of TrES-3b}

\correspondingauthor{Parijat Thakur}
\email{parijat@associates.iucaa.in, parijatthakur@yahoo.com}
\author{Vineet Kumar Mannaday}
\affil{Department of Pure and Applied Physics, Guru Ghasidas Vishwavidyalaya (A Central University), Bilaspur (C.G.) -495 009, India}
\author[0000-0002-0786-7307]{Parijat Thakur}
\affiliation{Department of Pure and Applied Physics, Guru Ghasidas Vishwavidyalaya (A Central University), Bilaspur (C.G.) -495 009, India}
\author{Ing-Guey Jiang}
\affiliation{Department of Physics and Institute of Astronomy, National Tsing-Hua University, Hsinchu, Taiwan}
\author{D. K. Sahu}
\affiliation{Indian Institute of Astrophysics, Bangalore -560 034, India}
\author{Y. C. Joshi}
\affiliation{Aryabhatta Research Institute of Observational Sciences (ARIES), Manora Peak, Nainital -263 002, India}
\author{A. K. Pandey}
\affiliation{Aryabhatta Research Institute of Observational Sciences (ARIES), Manora Peak, Nainital -263 002, India}
\author{Santosh Joshi}
\affiliation{Aryabhatta Research Institute of Observational Sciences (ARIES), Manora Peak, Nainital -263 002, India}
\author{Ram Kesh Yadav}
\affiliation{National Astronomical Research Institute of Thailand (NARIT), Sirindhorn AstroPark, 260 Moo 4, T. Donkaew, A. Maerim, Chiangmai, 50180 Thailand}
\author{Li-Hsin Su}
\affiliation{Department of Physics and Institute of Astronomy, National Tsing-Hua University, Hsinchu, Taiwan}
\author{Devesh P. Sariya}
\affiliation{Department of Physics and Institute of Astronomy, National Tsing-Hua University, Hsinchu, Taiwan}
\author{Li-Chin Yeh}
\affiliation{Institute of Computational and Modeling Science, National Tsing-Hua University, Hsinchu, Taiwan}
\author{Evgeny Griv}
\affiliation{Department of Physics, Ben-Gurion University, Beer-Sheva 84105, Israel}
\author{David Mkrtichian}
\affiliation{National Astronomical Research Institute of Thailand (NARIT), Sirindhorn AstroPark, 260 Moo 4, T. Donkaew, A. Maerim, Chiangmai, 50180 Thailand}
\author{Aleksey Shlyapnikov}
\affiliation{Crimean Astrophysical Observatory, 298409, Nauchny, Crimea}
\author{Vasily Moskvin}
\affiliation{Crimean Astrophysical Observatory, 298409, Nauchny, Crimea}
\author{Vladimir Ignatov}
\affiliation{Crimean Astrophysical Observatory, 298409, Nauchny, Crimea}
\author{M. Va\v{n}ko}
\affiliation{Astronomical Institute, Slovak Academy of Sciences, SK-059 60 Tatransk\'{a} Lomnica, Slovakia}
\author{\c{C.} P\"{u}sk\"{u}ll\"{u}}
\affiliation{Canakkale Onsekiz Mart University, Faculty of Sciences and Arts, Physics Department, 17100 Canakkale, Turkey}



\begin{abstract}
We present twelve new transit light curves of the hot-Jupiter TrES-3b observed during $2012-2018$ to probe the transit timing variation (TTV). By combining the mid-transit times determined from these twelve transit data with those re-estimated through uniform procedure from seventy one transit data available in the literature, we derive new linear ephemeris and  obtain the timing residuals that suggest the possibility of TTV in TrES-3 system. However, the frequency analysis shows that the possible TTV is unlikely to be periodic, indicating the absence of an additional body in this system. To explore the other possible origins of TTV, the orbital decay and apsidal precession ephemeris models are fitted to the transit time data. We find decay rate of TrES-3b to be $\dot{P_q}= -4.1 \pm 3.1$ $ms$ ${yr}^{-1}$ and the corresponding estimated modified tidal quality factor of ${Q}^{'}_{\ast}$ $\sim 1.11 \times {10}^{5}$ is consistent with the theoretically predicted values for the stars hosting the hot-Jupiters. The shift in the transit arrival time of TrES-3b after eleven years is expected to be ${T}_{shift}\sim 69.55 \ s$, which is consistent with the $RMS$ of the timing residuals. Besides, we find that the apsidal precession ephemeris model is statistically less probable than the other considered ephemeris models. It is also discussed that despite the linear ephemeris model appears to be the most plausible model to represent the transit time data, the possibility of the orbital decay cannot be completely ruled out in TrES-3 system. In order to confirm this, further high-precision and high-cadence follow-up observation of transits of TrES-3b would be important.  
\end{abstract}


\section{Introduction} \label{sec:intro}
Hot-Jupiters are short period ($P<10$ days) gas-giant Jupiter-like extra-solar planets, detected in tight orbits ($a<0.1$ AU) to their host stars. Since the discovery of first hot-Jupiter 51 Pegasi b \citep{Mayor95}, around a Sun-like star, more than four thousand  extra-solar planets\footnote{http://exoplanet.eu/catalog/} have been confirmed so far. Of these, 394 extra-solar planets in wide range of masses ($0.36 \ {M}_{J}\leq{M}_{P}\leq11.8 \ {M}_{J}$) are referred as  hot-Jupiters and majority of them are detected using transit method. The photometric study of these transiting hot-Jupiters are of vital importance.  Due to their short periods and strong transit signals, a long-term photometric follow-up observations of transits of these systems help in improving  the estimates of their physical and orbital parameters \citep[e.g.,][]{Sozz09,Montal12,Maci13a,Kund13,Colln17}. 

The improved estimate of mid-transit time from high-precision transit photometry allows to refine the transit ephemeris. The multi-epoch, high-precision transit photometry also provides an opportunity to  examine the transit timing variations (hereafter TTVs) of known planets, which could be due to presence of additional bodies in the planetary system when the TTV signal is periodic \citep{Mirald02,Agol05,Holm05,Heyl07,Jiang13,Jiang16,Maci15,Maci16,Misl15,Thakur18}. The close proximity of the massive hot-Jupiters to their host stars, makes them an ideal laboratory to test the long standing theoretical predictions of orbital decay and apsidal precession,  induced by the tidal interactions between hot-Jupiters and their host stars \citep[see][]{Ragoz09,Levrd09,Matsu10,Adam10,Maci16,Patra17,Csizma19}. These two phenomena are the other possible reasons to produce TTVs in the hot-Jupiter systems, which can be examined with the precise transit data if available for decade or more \citep{Maci16,Patra17}.
 
The orbital decay can be produced by the transfer of planet{'}s orbital angular momentum to star'{s} spin through the tidal dissipation \citep[e.g.,][]{Rasio96,Levrd09,Matsu10}, whereas the apsidal precession of non-zero eccentric orbits of hot-Jupiter systems can mainly be produced due to non-spherical mass components of gravitational quadruple fields created by tidal bulges raised on the planets \citep{Ragoz09,Maci16,Maci18,Patra17,Csizma19}. Probing these two phenomena in hot-Jupiter systems is considered to be very important, since the decay rate provides direct estimation of modified tidal quality factor $({Q}^{'}_{*})$ of host stars that indicates the efficiency of tidal dissipation within the host stars and remaining lifetime of hot-Jupiters \citep{Levrd09,Hellr09,Matsu10,Birkby14,Blecc14,Maci16,Patra17}. However, the apsidal precession rate provides direct estimation of planetary Love number $({k}_{p})$ that can be used to  infer interior density distribution of hot-Jupiters and also allows to confirm the presence or absence of massive cores in these planets \citep{Ragoz09}.

The tentative detection of decreasing period of some hot-Jupiters (OGLE-TR-113b: \cite{Adam10}; WASP-43b: \citet{Blecc14,Jiang16}; WASP-18b: \cite{Hellr09}; WASP-4b: \cite{Bouma19}) are still under debate,  as they could not be confirmed with the further observations (OGLE-TR-113b: \cite{Hoyer16a}; WASP-43b: \cite{Hoyer16b}; WASP-18b: \cite{Wilkn17}; WASP-4b: \cite{South19}). Recently, \cite{Maci16,Maci18} and \cite{Patra17} have reported decreasing period of WASP-12b that could be the first direct detection of orbital decay in any hot-Jupiter systems. However, they have also proposed for further follow-up observations of transit of WASP-12b, as the tidally induced orbital precession as an alternative scenario is still there to explain observed period shrinkage.

With a semi-major axis of $a = 0.0226 \ AU$, and planetary mass of ${M}_{p} = 1.92 \ {M}_{J}$, TrES-3b is one of the  close-in massive hot-Jupiters, which orbits around a G-type star (V = 12.4 mag) once in every 1.3 days \citep{Dono07}. Because of its strong transit signal and ultra-close proximity to the host star, this planetary system has been extensively followed-up for more than decade to improve the estimates of physical and orbital parameters, as well as to probe the possibility of additional planet through TTV analysis. For example, \cite{Sozz09} have performed both the radial velocity and photometric observations of TrES-3 system and reported the improved estimates of physical and orbital parameters. In order to search the distant massive companions to better understand the orbital evolutions of close-in hot-Jupiters, \cite{Knutson14} have performed the radial velocity observations for 51 hot-Jupiter systems including TrES-3 system. Although they have not found any evidence of an additional distant massive companion in TrES-3 system, a significant eccentricity of $e={0.17}^{+0.032}_{-0.031}$ was reported for TrES-3b. On the other hand, \cite{Bonomo17} have reanalyzed all the radial velocity data observed by \cite{Sozz09} and \cite{Knutson14} through a homogeneous procedure and stated that the eccentricity of TrES-3b is consistent with zero rather than a significant eccentricity found by \cite{Knutson14}. In addition to above, \cite{Sozz09},  \cite{Lee11}, \cite{Jiang13}, and \cite{Sun18} have proposed the presence of additional planet in TrES-3 system based on TTV analysis, whereas no evidence of additional planet was found by several authors \citep{Gibs09,Kund13,Vanko13,Pusk17,Ricc17}. Because of these contradictory findings and lack of strictly periodic TTV signal, nothing could be concluded regarding the presence of additional planet in this planetary system. However, most of previous authors have proposed to carry out further high-precision and high-cadence follow-up observations of this hot-Jupiter system to confirm their findings. Besides this, TrES-3b has been theoretically proposed to be a potential candidate to examine the orbital decay \citep[see][]{Levrd09,Matsu10,Penev18} and the apsidal precession \cite[see][]{Ragoz09}. Keeping this in mind, \cite{Sun18} have recently examined TrES-3 system and not found any indication of orbital decay. However, they have just used transit data only spanning over 4.5 years and not adopted a uniform procedure to calculate mid-transit times. As of now transit observations have been expanded over the decade, it would be worth to further explore the possible presence of additional planet, as well as the orbital decay and apsidal precession in TrES-3 system by including new transit observations to the transit data available in the literature for previous observed epochs.

In this paper, we present twelve new transit light curves of TrES-3b observed on different epochs. In order to perform the precise timing analysis for TrES-3 system, our newly observed transit light curves were combined with seventy one transit light curves available in the literature. Using the mid-transit times derived from these eighty three transit light curves with a uniform procedure, we examine the possibility of presence of additional planet, orbital decay, and apsidal precession in TrES-3 system. The remainder of the paper is organized as follows. In Section 2, we describe the details of our observations and data reduction procedure.  Section 3 presents the methodology used to analyze the transit light curves, as well as to derive the transit parameters. The estimation of new ephemeris and timing analysis of TrES-3 system are given in Section 4. The implications drawn from the linear, orbital decay and apsidal precession ephemeris models are discussed in Section 5. Finally the last section is devoted for the concluding remarks. 
\section{Observational Data} \label{sec:observation}
\subsection{Observations and Data Reduction}
The observations of twelve new transits of TrES-3b were carried out using  the 2-m Himalayan Chandra Telescope (HCT) at the  Indian Astronomical Observatory (IAO), Hanle, India, the 1.3-m Devasthal Fast Optical Telescope (DFOT) at the Aryabhatta Research Institute of Observational Sciences (ARIES), Nainital, India, and the 1.25-m AZT-11 telescope at the Crimean Astrophysical Observatory (CrAO) in Nauchny, Crimea. All the transit observation were made in R band, in order to minimize the effects of stellar limb-darkening and color dependent atmospheric extinction, as well as to achieve the high-cadence photometric observations \citep{Holman06}. The log of our observations is given in Table 1, whereas the specification of telescopes and CCD detectors used are listed in Table 2. As can be seen in Table 1, we have observed the transits of TrES-3b in focused (Run 1-2), slightly defocused (Run 3-5 and Run 7-12), and heavily defocused (Run 6) modes of the telescopes depending on mirror diameter, detector size and weather conditions. In order to avoid the saturation of CCD images with longer exposure time, we had to defocus the telescope heavily in case of the observing Run 6. However, the telescope defocusing technique allows to improve the precision of the photometric observation \citep[see][]{South09a,South09b,Hinse15,Maci15,Pusk17}.

All the science images of the TrES-3 system taken during each transit event were pre-processed using the standard tasks available within {\it IRAF} \footnote {IRAF (Image Reduction and Analysis Facility) is distributed by the National Optical Astronomy Observatories, which are operated by the Association of Universities for Research in Astronomy, Inc., under cooperative agreement with the National Science Foundation. For more details, http://iraf.noao.edu/} for trimming, bias-subtraction, and flat-fielding.  After the pre-processing, aperture photometry was performed on the TrES-3 and its nearby 2-8 comparison stars whose brightness and color are similar to those of TrES-3 using the {\it `daophot'} task within {\it IRAF}. The aperture size was allowed to vary in such a manner that it should give minimum scattering in the out-of-transit (OOT) data. This was usually 2-3 times the full width at half maximum (FWHM) of the stellar point spread function (PSF), whose range for each transit event is given in Table 1. In order to select the comparison stars, we followed the procedure given in \cite{Jiang16} and checked the correlations of out-of-transit (OOT) flux of TrES-3 with those of its nearby stars present in the same field by calculating the Pearson's correlation coefficient, $r$. The stars with $r > 0.90$ were chosen as the comparison stars. These strong correlations show the brightness consistency between the TrES-3 and the chosen comparison stars. For each transit event, several transit light curves were obtained by dividing the flux of TrES-3 by the flux of each comparison star. Further, the light curves were also obtained by dividing the sum of fluxes from the different combinations of the comparison stars to the flux of TrES-3 \citep[see][]{Gibs09,Jiang16,Patra17}. It is also ensured here that the OOT flux variations should not correlate with the airmass, indicating the similar colors of TrES-3 and objects represented with the different combinations of comparison stars. The light curve of each transit event is finalized by identifying the object with the best combination of comparison stars that produces minimum root mean square ($RMS$) in the OOT data \citep[see][]{Hoyer16b,Turner17}. The number of comparison stars used to obtain each transit light curve and corresponding OOT $RMS$ are also listed in Table 1. It is worth mentioning here that the used aperture size and number of comparison stars are different in each transit event, which may be due to varied sky conditions and change in field orientation \citep[see][]{Gibs09,Colln17}. To remove time-varying atmospheric effects, the transit light curves were normalized by fitting a linear function to OOT data, which leads to OOT flux close to unity. The time stamps as Heliocentric Julian Days (HJD) in the transit light curves were converted to Barycentric Julian Days (BJD) with the time standard Barycentric Dynamical Time (TDB), {\it i.e.}, TDB-based BJD, using online tool provided by \citet{East10}\footnote{http://astroutils.astronomy.ohio-state.edu/time/hjd2bjd.html}. The normalized transit light curves of TrES-3 system obtained from our observations, along with their best fit-models and residuals are shown in Figure 1 (see Section 3 for details).

\begin{table*}
\begin{center}
\caption{Log of Observations}
\label{tab:1}
\small\addtolength{\tabcolsep}{-2pt}
\begin{tabular}{lccccccccc}
\toprule
Run & UT Date  & Telescope & ${Mode}^{a}$ & Interval (HJD-2450000) & ${ExpT}^{b}$  & No. & ${ Range}^{c}$ & ${No.}^{d}$ & ${OOT \ RMS}^{e}$ \\ 
& & & of & & & of & of & of & \\
& & & Tel. &  &  & images & FWHM &   CS & \\
\hline
1 	& 29.05.2012 & 1.3-m DFOT 		&  F & 6077.24390 - 6077.33862 & 150 & 42 & 3-4 & 6	& 0.16\\
2 	& 20.06.2012 & 1.25-m AZT-11	&  F & 	6099.42460 - 6099.50950 & 30 & 237 & 2-3 & 2 & 7.74\\
3 	& 10.04.2013 & 1.3-m DFOT		& SDF &   6393.30567 - 6393.46768 & 120 & 89 & 5-7& 2 & 1.63\\
4 	& 10.05.2013 & 1.3-m DFOT		& SDF &   6423.28917 - 6423.46973 & 120 & 66 & 4-6 & 4	& 1.40\\
5 	& 18.05.2013 & 1.3-m DFOT		& SDF &   6431.19286 - 6431.37571 & 120 & 150 & 4-6 &  3 & 0.60\\
6 	& 30.03.2014 & 1.3-m DFOT		& HDF &   6747.30530 - 3647.42750 & 120 	& 79 & 20-23 &  5 & 0.17\\
7 	& 07.03.2018 & 2-m HCT			& SDF &  	8185.38013 - 8185.49678 & 45-60  & 57 & 9-13 & 4	& 0.58\\
8 	& 11.03.2018 & 2-m HCT  		& SDF &   8189.32393 - 8189.43049 & 30-60  & 58 & 6-9 & 2	& 2.13\\
9 	& 24.03.2018 & 2-m HCT  		& SDF &   8202.35153 - 8202.46584 & 120-60  & 42 & 6-8 & 5	& 0.91\\
10 	& 28.03.2018 & 2-m HCT 			& SDF &   8206.28081 - 8206.41188 & 30-60  & 68 & 8-10 & 3	& 1.7\\
11 	& 10.04.2018 & 2-m HCT 			& SDF &   8219.34240 - 8219.46581 & 45-60  & 42 & 6-9 & 7	& 0.99\\
12 	& 14.04.2018 & 2-m HCT 			& SDF &   8223.24881 - 8223.39519 & 30-60  & 73 & 6-8 & 3	& 2.49\\
\hline
\end{tabular}
\end{center}
{Notes:\\ 
$^a$ {Mode of the telescope during the transit observation: F indicates focused mode, SDF indicates slightly defocused mode, and HDF indicates heavily defocused mode}\\
$^b$ Exposure time in units of second\\
$^c$ Range of the FWHM of the stellar point spread function (PSF) in units of pixels \\
$^d$ Number of comparison stars\\
$^e$ Out-of-transit root mean square in units of ${10}^{-3}$}
\end{table*}

\begin{table*}
\begin{center}
\caption{Specification of Telescopes and CCD detectors used in This Work}
\label{tab:2}
\begin{tabular}{ccccccc}
\toprule
Telescope and CCD detector & CCD size & Field of View & Plate Scale &  Readout Noise & Gain \\ 
&	&	(arcmin$\times$arcmin)	&	(arcsec ${pixel}^{-1}$)	& (${e}^{-}$)	& (${e}^{-}/ADU$)\\
\hline
2-m HCT, SITe CCD & $2K\times2K$ & $10\times10$ & 0.296 & 4.8 & 1.22 \\
1.3-m DFOT, Andor CCD & $2K\times2K$ & $18\times18$ & 0.54  & 7.0 & 2.0 \\
1.25-m AZT-11, ProLine PL230 & $2K\times2K$ & $10.9\times10.9$ & 0.32  & 12.9 & 1.94 \\
\hline
\end{tabular}
\end{center}
\end{table*}

\subsection{Other Observational Data from Literature}
In addition to our twelve new transit observations, we have also taken seventy one transit light curves from the literature. These include eight transit light curves from \citet{Sozz09}, nine from \citet{Gibs09}, one from \citet{Coln10}, four from \citet{Lee11}, five from \citet{Jiang13}, ten from \citet{Kund13}, seven from \citet{Turnr13}, eleven from \citet{Vanko13}, five from \citet{Ricc17}, and eleven from \citet{Pusk17}. In total, eighty three transit light curves of TrES-3 system spanning over  more than a decade,  are included in this work.

\section{Light Curve Analysis}
To determine the physical and orbital parameters of TrES-3 system from our twelve new transit light curves, the Transit Analysis Package (TAP) described by \citet{Gaz12} was utilized. The TAP uses Markov Chain Monte Carlo (MCMC) technique to fit the observed transit light curves with the model light curves of \citet{Mandagol02} derived from a simple two-body star-planet system. In order to take into account the effect of limb-darkening across the stellar disk, a quadratic limb-darkening law \citep{Kopal50} is also implemented in TAP. As the photometric time series may be affected by both the temporally uncorrelated (white) and temporally correlated (red) noises, the wavelet-based likelihood technique of \citet{Cartwin09} is employed in TAP to robustly estimate parameter uncertainties. For more details description of TAP and wavelet-based likelihood techniques, we simply refer the  readers to  \citet{Cartwin09}, \citet{Fult11}, and \citet{Gaz12}.

In order to set up the initial values of parameters, as well as to analyze the transit light curves, we followed the same procedure as adopted by \cite{Jiang13}. The ratio of planet to star radius ({R$_p$/R$_\ast$}) and mid-transit time (${T}_{m}$) were treated as free parameters in the light curve analysis. However, the eccentricity of orbit (${e}$) and longitude of pariastron ({$\omega$}) were set to zero as suggested by \citet{Dono07} and \citet{Fres10}, and the orbital period ($P$) was kept fixed to the same value as given in \cite{Sozz09}. The remaining parameters, namely, ratio of semi-major axis to stellar radius ({a/R$_\ast$}), orbital inclination ($\it {i}$), linear (u$_1$) and quadratic (u$_2$) limb-darkening coefficients were fitted under Gaussian penalties by adopting the same procedure as given in \citet{Jiang13}. Moreover, the initial values of the parameters {a/R$_\ast$}, {\it i}, and {R$_p$/R$_\ast$} were adopted from \citet{Sozz09}.  

For the filters $U$, $B$, $V$, $R$, $I$, and Sloan $i$, $g$, $r$, $z$, $u$, the initial values of limb-darkening coefficients u$_1$ and u$_2$ were linearly interpolated from the tables of \citet{Clar00,Clar04} using the JKTLD\footnote{JKTLD code is available from http://www.astro.keele.ac.uk/∼jkt/codes.html} code \citep{South15} with the stellar parameters such as effective temperature $({T}_{eff}=5650.0 \ K)$, stellar surface gravity $({\log g=4.40} \ cm \ {s}^{-2})$, metallicity $([Fe/H]=-0.19$), and micro-turbulence velocity $({V}_{t}=2.0 \ km \ {s}^{-1})$ taken as in \cite{Sozz09}. The details of initial parameter setting for our light curve analysis are given in Table 3. 

In addition to eight transit light curves in  $R$ filter and one in $I$ filter observed by \citet{Vanko13}, we considered their two more transit light curves that were observed in clear and Luminance filters. Since the clear filter covers $V$ and $R$ bands \citep{Maci13b}, the limb darkening coefficients u$_1$ and u$_2$ for clear filter are taken as the average of their value in  $V$ and $R$ filters. However, the limb-darkening coefficients derived in  $V$ filter were taken for the Luminance filter. The limb-darkening coefficients derived in Sloan $r$ filter were used for analysis of ten transit light curves of \citet{Kund13}. Moreover, the values of limb-darkening coefficients reported in \citet{Turnr13} for their seven transit light curves observed in Harris $B$, $V$, and $R$ filters were directly adopted from their paper. Table 4 lists all theoretical values of limb-darkening coefficients for different filters considered in this work. As in \citet{Jiang13}, these values of limb-darkening coefficients were taken as  initial values and fitted under Gaussian penalties with $\sigma=0.05$, to consider the possible small differences between best-fitted limb-darkening coefficients and those interpolated from the tables of \citet{Clar00, Clar04}. 

For each transit light curve analysis, we used five MCMC chains with lengths of ${10}^{6}$ links each. To obtain the well sampled posterior probability distribution, we specified the desired acceptance rate of $\sim 0.44$ for each of model parameters ${T}_{m}$, ${\it i}$, {a/R$_\ast$}, {R$_p$/R$_\ast$}, u$_1$, and u$_2$ \citep[see][]{Ford06,Gaz12}. After this, the TAP automatically designs and updates the characteristic size of model parameter jump between links \citep[$\beta$, as defined in][]{Ford06,Gaz12} and continues this process until the desired acceptance rates are achieved for model parameters. The set of $\beta$ values obtained corresponding to desired acceptance rates are locked in and then efficient calculation of MCMC chain begins \citep[see][]{Gaz12}. In order to test for non-convergence of MCMC chains, the TAP employs the Gelman-Rubin statistics (hereafter G-R statistics) and analyzes the likelihood that multiple chains have converged to the same parameter space \citep{Gelman03,Ford06,Gaz12}. In this analysis, it calculates G-R statistics \citep[${\hat{R}} (z)$, as defined in][]{Ford06} for each model parameter, as well as also estimates the effective number of independent samples \citep[${\hat{T}} (z)$, as defined in][]{Ford06}. This process continues by automatically extending the chains until ${\hat{R}} (z) \leq 1.01$  and ${\hat{T}} (z) \geq 1000$ \citep[see][]{Ford06}. When all these tests based on ${\hat{R}} (z)$ and ${\hat{T}} (z)$ are satisfied, it is considered that the calculated MCMC chains have sufficiently mixed and achieved a state of convergence \citep[see][]{Ford06}. Once all the MCMC chains have converged, the TAP automatically discards the first 10\% or 10,000 (whichever is greater) links from each chain to reduce the effect of initial parameter values and adds the remaining chains together for Bayesian parameter extraction. The 50.0 percentile level (median), as well as the 15.9 and 84.1 percentile levels (i.e., 68\% credible intervals) of the posterior probability distribution for each model parameter are considered as the best-fit value, as well as its lower and upper $1\sigma$ uncertainties, respectively. For our twelve new transit light curves, the best-fit values of the parameters ${T}_{m}$, ${\it i}$, {a/R$_\ast$}, {R$_p$/R$_\ast$}, u$_1$ and u$_2$ along with their $1\sigma$ uncertainties are listed in Table 5. The first transit of TrES-3b shown in \cite{Sozz09} was defined to be epoch E=0, and other transit epochs considered in this work were calculated accordingly. The transit light curves obtained from our twelve new observations with their best fit models and corresponding residuals are shown in Figure 1. As we followed the procedure adopted by \citet{Jiang13} for transit light curve analysis using TAP, all the twenty three mid-transit times and their $1\sigma$ uncertainties reported for TrES-3b in their paper were directly used for this study. In order to maintain the homogeneity in transit light curve modeling and fitting procedure for precise TTV analysis, the mid-transit times and their $1\sigma$ uncertainties for other forty eight transit light curves of TrES-3b taken from literature were re-determined individually using TAP by employing the same procedure as given in \cite{Jiang13}. The mid-transit times (${T}_{m}$) along with their $ 1\sigma$ uncertainties and corresponding epochs (E) for the total number of eighty three transit light curves used in this paper are gathered in Table 6.

\begin{table*}
\begin{center}
\caption{The Initial Parameter Setting}
\label{tab:2}
\begin{tabular}{lcc}
\toprule
Parameter & Initial Value & During MCMC Chains\\
\hline
$P$ (days) & 1.30618581 & Fixed\\
${\it i}$ (deg) & 81.85 & A Gaussian prior with $\sigma$ = 0.16\\ 
{a/R$_\ast$} & 5.926 & A Gaussian prior with $\sigma$ = 0.056\\
{R$_p$/R$_\ast$} & 0.1655 & Free \\
${T}_{m}$ & Set by eye & Free\\
u$_1$ & \citet{Clar00, Clar04} & A Gaussian prior with $\sigma$ = 0.05\\
u$_2$ & \citet{Clar00, Clar04} & A Gaussian prior with $\sigma$ = 0.05\\
\hline
\end{tabular}\\
\end{center}
{Note: 
The initial values of $P$, ${\it i}$, {a/R$_\ast$}, and {R$_p$/R$_\ast$} are set as the values in \citet{Sozz09}.}
\end{table*}

\begin{table*}
\begin{center}
\caption {The Theoretical Limb-darkening Coefficients for TrES-3 Star}
\label{tab:3}
\begin{tabular}{lcc}
\hline
Filter & u$_1$ & u$_2$\\
\hline
U$^{a}$ & 0.8150	& 0.0490\\
B$^{a}$ & 0.6379& 0.1792\\
V$^{a}$ & 0.4378 & 0.2933\\
R$^{a}$ & 0.3404 & 0.3190\\
I$^{a}$ & 0.2576 & 0.3186\\
Sloan \ u$^{a}$ & 0.8112 & 0.0554\\
Sloan \ g$^{a}$ & 0.5535 & 0.2351\\
Sloan \ r$^{a}$ & 0.3643 & 0.3178\\
Sloan \ i$^{a}$ & 0.2777 & 0.3191\\
Sloan \ z$^{a}$ & 0.2179& 0.3162\\
Harris \ B$^{b}$ & 0.63712 & 0.17994\\
Harris \ V$^{b}$ & 0.43880 & 0.29264\\
Harris \ R$^{b}$ & 0.34156 & 0.31818\\ 
Clear$^{c}$ & 0.3891 & 0.30615\\
\hline
\end{tabular}\\
\end{center}
{Notes: \\
$^a$ Calculated for ${T}_{eff}=5650 K$, $\log{g}=4.40$ \ cm  ${s}^{-2}$, $[Fe/H]=-0.19$, and ${V}_{t}=2 \ km \ {s}^{-1}$.\\
$^b$  u$_1$ and u$_2$ directly adopted from \citet{Turnr13}. \\
$^c$ Calculated as the average of their value in  V and R filters.}
\end{table*}

\begin{figure*}
	\includegraphics[width=\columnwidth]{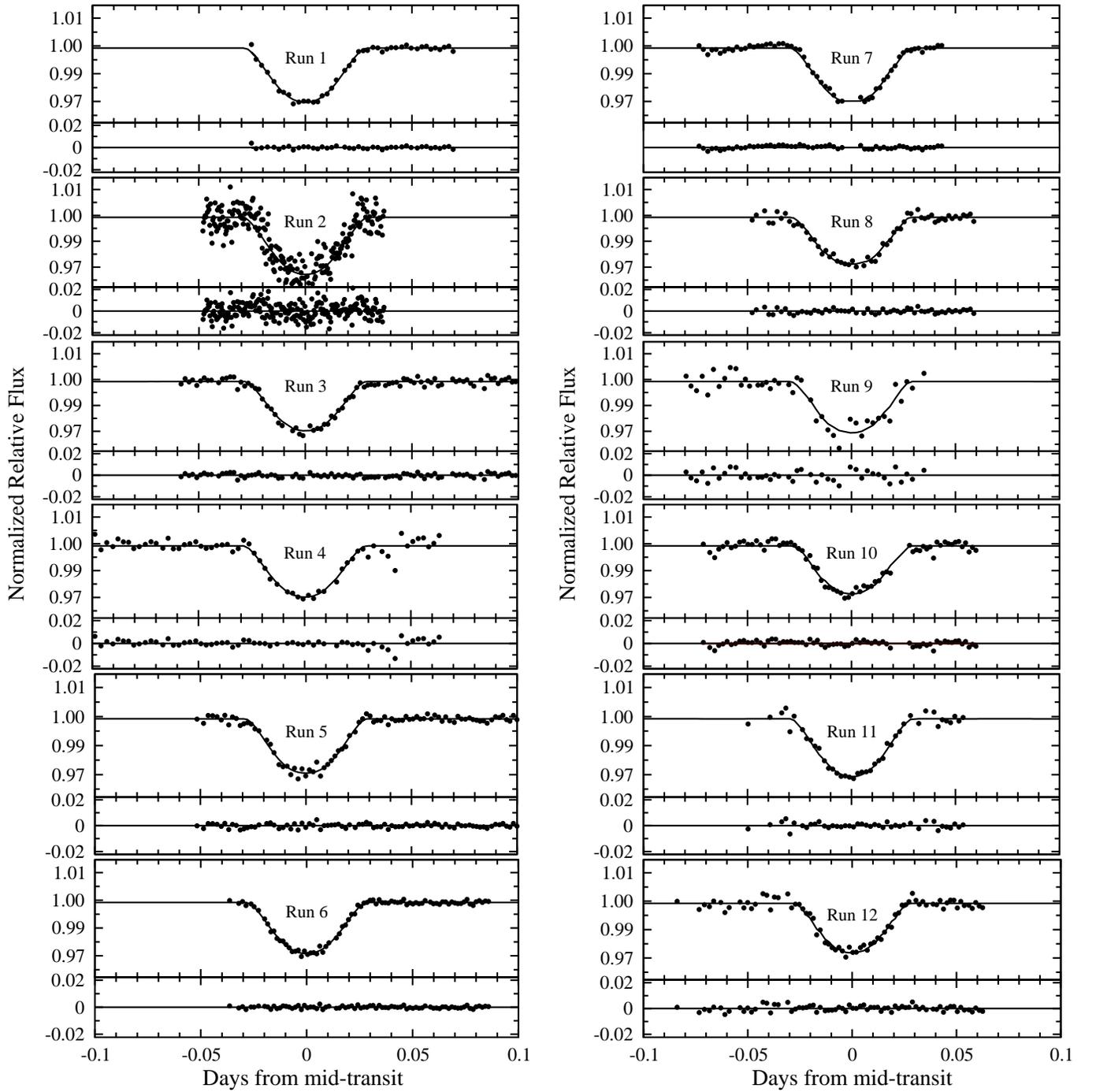}
    \caption{Normalized relative flux as a function of the time (the offset from mid-transit time and in TDB-based BJD) of twelve transit light curves of this work: points are the data and solid lines are best-fit models. The corresponding residuals  are shown at the bottom of the figures.}
    \label{fig:1}
\end{figure*}

\begin{table*}
	\centering
	\caption{The  Best-fit Values of Parameters ${T}_{m}$, ${\it i}$, {a/R$_\ast$},{R$_p$/R$_\ast$},  u$_1$, and u$_2$ for Twelve New Transit Light Curves }
	\label{tab:6}
	\small\addtolength{\tabcolsep}{-1.5pt}
	\begin{tabular}{cccccccc} 
		\hline
		Run & Epoch (E) & ${T}_{m}$ (in ${BJD}_{TDB}$) & ${\it i}$ (in deg) & {a/R$_\ast$} & {R$_p$/R$_\ast$} & u$_1$ & u$_2$\\
		\hline 
		1 & 1448 & $2456077.27003^{+0.00035}_{-0.00037}$ & $81.75^{+0.13}_{-0.13}$ & $ {5.967^{+0.047}_{-0.047}}$ & ${0.1734^{+0.0062}_{-0.0064}}$ & ${0.343^{+0.050}_{-0.050}}$ & $ {0.323^{+0.050}_{-0.050}}$\\
		\\[-0.6em]
		2 &1465 & $2456099.473370^{+0.0016}_{-0.0016}$ & $81.81^{+0.14}_{-0.14}$ & $ {5.944^{+0.054}_{-0.054}}$ & ${0.1830^{+0.011}_{-0.013}}$ & ${0.341^{+0.050}_{-0.050}}$ & ${0.322^{+0.050}_{-0.050}}$\\
		\\[-0.6em]
		3 & 1690 & $2456393.36471^{+0.00048}_{-0.00050}$ & $81.78^{+0.13}_{-0.14}$ & ${5.944^{+0.049}_{-0.050}}$ & ${0.1687^{+0.0066}_{-0.0059}}$ & ${0.344^{+0.050}_{-0.050}}$ & ${0.304^{+0.050}_{-0.050}}$\\
		\\[-0.6em]
		4 & 1713 & $2456423.40717^{+0.00079}_{-0.00079}$ & $81.83^{+0.14}_{-0.14}$ & ${5.933^{+0.052}_{-0.052}}$ & ${0.1674^{+0.0082}_{-0.0072}}$ & ${0.341^{+0.050}_{-0.050}}$ & $ {0.319^{+0.050}_{-0.050}}$\\
		\\[-0.6em]
		5 & 1719 & $2456431.24526^{+0.00032}_{-0.00031}$ & $81.87^{+0.12}_{-0.12}$ & $ {5.921^{+0.046}_{-0.046}}$ & ${0.1641^{+0.0042}_{-0.0034}}$ & ${0.340^{+0.049}_{-0.049}}$ & $ {0.319^{+0.049}_{-0.049}}$\\
		\\[-0.6em]
		6 & 1961 & $2456747.34245^{+0.00019}_{-0.00019}$ & $81.78^{+0.11}_{-0.12}$ & $ {5.955^{+0.043}_{-0.043}}$ & ${0.1665^{+0.0037}_{-0.0032}}$ & ${0.341^{+0.049}_{-0.049}}$ & $ {0.321^{+0.049}_{-0.050}}$\\
		\\[-0.6em]	
		7 & 3062 & $2458185.45419^{+0.00060}_{-0.00064}$ & $81.81^{+0.14}_{-0.14}$ & ${5.944^{+0.051}_{-0.051}}$ & ${0.1714^{+0.0083}_{-0.0074}}$ & ${0.342^{+0.050}_{-0.050}}$ & $ {0.321^{+0.050}_{-0.050}}$\\
		\\[-0.6em]
		8 & 3065 & $2458189.37246^{+0.00068}_{-0.00065}$ & $81.88^{+0.14}_{-0.14}$ & ${5.915^{+0.052}_{-0.051}}$ & ${0.1609^{+0.0078}_{-0.0065}}$ & ${0.339^{+0.050}_{-0.049}}$ & ${0.317^{+0.049}_{-0.049}}$\\
		\\[-0.6em]
		9 & 3075 & $2458202.43178^{+0.0015}_{-0.0015}$ & $81.88^{+0.16}_{-0.16}$ & ${5.921^{+0.055}_{-0.055}}$ & ${0.1690^{+0.012}_{-0.012}}$ & ${0.338^{+0.050}_{-0.050}}$ & ${0.317^{+0.050}_{-0.050}}$\\
		\\[-0.6em]
		10 & 3078 & $2458206.35282^{+0.00076}_{-0.00071}$ & $81.82^{+0.13}_{-0.14}$ & ${5.934^{+0.051}_{-0.051}}$ & ${0.1616^{+0.0072}_{-0.0063}}$ & ${0.342^{+0.049}_{-0.049}}$ & $ {0.321^{+0.050}_{-0.050}}$\\
		\\[-0.6em]
		11 & 3088 & $2458219.41292^{+0.00057}_{-0.00058}$ & $81.79^{+0.14}_{-0.14}$ & ${5.942^{+0.051}_{-0.051}}$ & ${0.1731^{+0.0081}_{-0.0070}}$ & ${0.343^{+0.050}_{-0.050}}$ & ${0.322^{+0.050}_{-0.050}}$\\
		\\[-0.6em]
		12 & 3091 & $2458223.33346^{+0.00053}_{-0.00050}$ & $81.82^{+0.13}_{-0.13}$ & ${5.941^{+0.050}_{-0.050}}$ & ${0.1584^{+0.0064}_{-0.0054}}$ & ${0.341^{+0.050}_{-0.050}}$ & ${0.320^{+0.050}_{-0.050}}$\\
		\hline 
	\end{tabular}
\end{table*}

\section{Transit Timing Analysis}
\subsection{New Ephemeris}
We derived new ephemeris for orbital period $P$ and mid-transit time ${T}_{0}$ of TrES-3b by fitting a linear ephemeris model,
\begin{equation}
T^{c}_{m} (E) = T_0 + EP,
\end{equation}
to the eighty three mid-transit times ${T}_{m}$ as a function of epoch E given in Table 6 using the $emcee$ MCMC sampler implementation \citep{Foreman13}, where ${T}^{c}_{m}$, E, $P$, and ${T}_{0}$ are the calculated mid-transit time, epoch, orbital period, and mid-transit time at E = 0, respectively. In order to estimate the new linear ephemeris for orbital period $P$ and mid-transit time ${T}_{0}$ using MCMC technique, we assumed a Gaussian likelihood and imposed uniform priors on the parameters $P$ and ${T}_{0}$. The uniform prior used for each parameter is listed in Table 7. We used 100 walkers and then ran 300 steps of every walker as an initial burn-in to adjust the step size for each parameter \citep[e.g.,][]{Garhart18}. Considering the final position of the walkers after the 300 steps as the initial position, we further ran 20,000 steps per walker of MCMC to determine the best-fit parameters of linear ephemeris model and their uncertainties \citep[e.g.,][]{Hoyer16a,Hoyer16b}. We ensured here the efficient calculation of well sampled MCMC chain, since the estimated mean acceptance fraction of $\sim 0.44$ was found to be consistent within the ideal range of 0.2-0.5 \citep[see][]{Foreman13,Cloutier16,Stefansson17}. To assess the convergence of MCMC chain, we estimated the integrated autocorrelation time of the chain averaged across parameters \citep{Goodman10,Foreman13} and found its value to be $\sim 19$ steps. This suggests that only after $\sim 19$ steps, the drawn samples of model parameters become independent and start converging toward the reasonable parameter space \citep{Hogg18}. By dividing the estimated value of integrated autocorrelation time of $\sim 19$ steps to 20,000 steps per walker of MCMC, we determined the effective number of independent samples to be  $\sim 1052$ \citep[see][]{Mede17}, which was found to be larger than its minimum threshold value of $50$ per walker set in our MCMC analysis as suggested by emcee group\footnote{http://emcee.readthedocs.io/en/sable/tutorials/autocorr/}\citep[see][]{Shinn19}. From our MCMC analysis, the above obtained characteristics of the acceptance fraction and the effective number of independent samples calculated using the estimated value of integrated autocorrelation time confirm their acceptability and reliability. This also supports the well performance and convergence of MCMC chain.  

In order to avoid the strongly correlated parameters, the initial 37 steps (i.e., nearly two times the estimated value of integrated autocorrelation time) were also discarded as a final burn-in from the 20,000 steps per walker of MCMC \citep[see][]{Almenara16,David18}. Finally, the remaining samples of model parameters $P$ and $T_0$ were used for Bayesian parameter extraction. The 50.0 percentile level (median) of the posterior probability distribution for each model parameter is inferred as the best-fit value, while the 16.0 and 84.0 percentile levels (i.e., 68\% credible intervals) of the posterior probability distribution are considered as its lower and upper $1\sigma$ uncertainties, respectively. 

The corner plot depicting the marginalized 1-D and 2-D posterior probability distributions for the parameters of linear ephemeris model is shown in Figure 2. In this plot, the 1-D histogram along the diagonal panel shows the posterior probability distribution for each parameter obtained by marginalizing over the other parameter, where the three vertical dashed lines represent the median and 68\% credible intervals. The off-diagonal panel shows the marginalized 2-D projection of posterior probability distribution for the covariation between the pair of parameters, with $1\sigma$, $2\sigma$ and $3\sigma$ contours. Moreover, the solid vertical and horizontal blue lines in off-diagonal panel show the best-fit model parameters. The symmetric and Gaussian-like posterior probability distribution of each model parameter (diagonal panel), as well as its best fit value lying within the smooth and smaller in size $1\sigma$ contour (off-diagonal panel) indicate the robust fitting of the linear ephemeris model to the transit time data with MCMC technique. This confirms the reliable estimation of the new linear ephemeris for orbital period $P$ and mid-transit time $T_0$ along with their $1\sigma$ uncertainties and are given in Table 7. The derived values of new ephemeris are consistent with those reported previously in the literature. The value of minimum ${\chi}^{2}$ of this model fit is $150.58$. Since the degree of freedom is $81$, the ${\chi}^{2}_{red} (81)$ is found to be $1.859$. The Bayesian Information Criterion ($BIC= {\chi}^{2}$ + k $\log{N}$, where k is the number of free parameters and N is the number of data points) corresponding to the best-fit is $159.41$. Using the new ephemeris, the timing residuals, (O-C), defined as the difference between observed mid-transit times, $T_m$ and the calculated mid-transit times, ${T}^{c}_{m}$, were calculated for each epoch E considered in this work and also given in Table 6. The timing residual as a function of epoch E is shown in Figure 3 and its $RMS$ is found to be $ \sim 69.86 \ s$. Since the linear ephemeris model (i.e., null-TTV model) provides a poor fit to the transit time data with ${\chi}^{2}_{red}>1$, there may be possibility of the TTV in TrES-3 system. 

\begin{center}
\small\addtolength{\tabcolsep}{-3pt}
\begin{longtable*}{cp{4.5cm}cp{4.5cm}cp{3cm}l}
\caption{Mid-transit Times $({T}_{m})$ and Timing Residuals \ \ \ (O-C) for Eighty Three Transit Light Curves} \label{tab:long}\\
\hline
\hline \multicolumn{1}{c}{\textbf{Epoch}} & \multicolumn{1}{c}{\textbf{${T}_{m}$ }} & \multicolumn{1}{c}{\textbf{O-C} } & \multicolumn{1}{l}{\textbf{Data Sources}}\\
(E)	&	\ \ \ \ \ \ \ \ \ \ $({BJD}_{TDB})$	& (days)	& \\
\hline
\endfirsthead
\multicolumn{4}{c}
{{\bfseries \tablename\ \thetable{} -- continued}} \\
\hline
\hline \multicolumn{1}{c}{\textbf{Epoch}} & \multicolumn{1}{c}{\textbf{${T}_{m}$}} & \multicolumn{1}{c}{\textbf{O-C}} & \multicolumn{1}{l}{\textbf{Data Sources}}\\
(E)	&	\ \ \ \ \ \ \ \ \ \ $({BJD}_{TDB})$	& (days)	& \\
\hline
\endhead
 0 & $2454185.91110^{+0.00020}_{-0.00020}$ & -0.0001245 & \citet{Sozz09}$^{a}$ \\
10 & $2454198.97359^{+0.00057}_{-0.00066}$ & 0.0005037 & \citet{Sozz09}$^{a}$ \\
22 & $2454214.64695^{+0.00032}_{-0.00036}$ & -0.0003703 & \citet{Sozz09}$^{a}$\\
23 & $2454215.95288^{+0.00033}_{-0.00031}$ & -0.0006265 & \citet{Sozz09}$^{a}$ \\
267 & $2454534.66317^{+0.00019}_{-0.00019}$ & 0.0002374 & \citet{Gibs09}$^{a}$ \\
268 & $2454535.96903^{+0.00039}_{-0.00037}$ & -0.0000887 & \citet{Sozz09}$^{a}$\\
281 & $2454552.94962^{+0.00020}_{-0.00022}$ & 0.0000810 & \citet{Sozz09}$^{a}$ \\
294 & $2454569.92982^{+0.00039}_{-0.00040}$ & -0.0001392 & \citet{Sozz09}$^{a}$ \\
313 & $2454594.74682^{+0.00037}_{-0.00034}$ & -0.0006765 & \citet{Sozz09}$^{a}$ \\
329 & $2454615.64621^{+0.00020}_{-0.00021}$ & -0.0002653 & \citet{Gibs09}$^{a}$ \\
342 & $2454632.62690^{+0.00020}_{-0.00019}$ & -0.0000045 & \citet{Gibs09}$^{a}$\\
355 & $2454649.60712^{+0.00019}_{-0.00017}$ & -0.0001957 & \citet{Gibs09}$^{a}$\\
358 & $2454653.52661^{+0.00091}_{-0.00092}$ & 0.0007357 & \citet{Gibs09}$^{a}$ \\
365 & $2454662.66984^{+0.00059}_{-0.00060}$ & 0.0006625 & \citet{Gibs09}$^{a}$ \\
371 & $2454670.50709^{+0.00034}_{-0.00034}$ & 0.0007955 & \citet{Gibs09}$^{a}$ \\
374 & $2454674.42521^{+0.00028}_{-0.00028}$ & 0.0003570 & \citet{Gibs09}$^{a}$ \\
381 & $2454683.56812^{+0.00042}_{-0.00041}$ & -0.0000362 & \citet{Gibs09}$^{a}$ \\
592 & $2454959.17120^{+0.0011}_{-0.0011}$ & -0.0022386 & \citet{Lee11}\\
596 & $2454964.40014^{+0.00088}_{-0.00095}$ & 0.0019567 & \citet{Vanko13} \\
597 & $2454965.70470^{+0.00023}_{-0.00021}$ & 0.0003305 & \citet{Kund13}\\
606 & $2454977.46000^{+0.0015}_{-0.0015}$ & -0.0000450 & \citet{Vanko13} \\
620 & $2454995.74657^{+0.00016}_{-0.00017}$ & -0.0000814 & \citet{Kund13}\\
620 & $2454995.74737^{+0.00040}_{-0.00044}$ & 0.0007186 & \citet{Turnr13}\\
627 & $2455004.88970^{+0.00018}_{-0.00018}$ & -0.0002546 & \citet{Turnr13}\\
637 & $2455017.95161^{+0.00033}_{-0.00030}$ & -0.0002064 & \citet{Turnr13}\\
658 & $2455045.38085^{+0.00060}_{-0.00063}$ & -0.0008760 & \citet{Vanko13} \\
665 & $2455054.52523^{+0.00018}_{-0.00017}$ & 0.0002008 & \citet{Coln10}\\
668 & $2455058.44480^{+0.0010}_{-0.0010}$ & 0.0012123 & \citet{Vanko13} \\
836 & $2455277.88206^{+0.00038}_{-0.00038}$ & -0.0008047 & \citet{Kund13}\\
849 & $2455294.86465^{+0.00039}_{-0.00038}$ & 0.0013651 & \citet{Lee11} \\
864 & $2455314.45500^{+0.00068}_{-0.00072}$ & -0.0010775 & \citet{Vanko13} \\
878 & $2455332.74259^{+0.00031}_{-0.00029}$ & -0.0000939 & \citet{Kund13}\\
885 & $2455341.88380^{+0.0011}_{-0.0010}$ & -0.0002187 & \citet{Jiang13}$^{a}$\\
898 & $2455358.86606^{+0.00076}_{-0.00074}$ & -0.0003474 & \citet{Jiang13}$^{a}$\\
898 & $2455358.86723^{+0.00068}_{-0.00070}$ & 0.0008226 & \citet{Lee11} \\
901 & $2455362.78470^{+0.0011}_{-0.00098}$ & -0.0002659  & \citet{Jiang13}$^{a}$\\
901 & $2455362.78568^{+0.00057}_{-0.00056}$ & 0.0007141 & \citet{Lee11}\\
904 & $2455366.70215^{+0.00080}_{-0.00077}$ & -0.0013744 & \citet{Jiang13}$^{a}$\\
911 & $2455375.84617^{+0.00089}_{-0.00090}$ & -0.0006576 & \citet{Jiang13}$^{a}$\\
913 & $2455378.45955^{+0.00090}_{-0.00084}$ & 0.0003500 & \citet{Vanko13}\\
942 & $2455416.33972^{+0.00056}_{-0.00056}$ & 0.0011210 & \citet{Vanko13}\\
952 & $2455429.39997^{+0.00046}_{-0.00045}$  & -0.0004907 & \citet{Vanko13}\\
965 & $2455446.38075^{+0.00021}_{-0.00019}$ & -0.0001309 & \citet{Vanko13}\\
992 & $2455481.64795^{+0.00018}_{-0.00018}$ & 0.0000424 & \citet{Kund13}\\
1116 & $2455643.61454^{+0.00034}_{-0.00034}$ & -0.0004530 & \citet{Vanko13}\\
1117 & $2455644.92122^{+0.00019}_{-0.00018}$ & 0.0000409 & \citet{Kund13}\\
1143 & $2455678.88252^{+0.00030}_{-0.00032}$ & 0.0005004 & \citet{Kund13}\\
1156 & $2455695.86223^{+0.00072}_{-0.00069}$ & -0.0002099 & \citet{Kund13}\\
1185 & $2455733.74164^{+0.00035}_{-0.00035}$ & -0.0001989 & \citet{Kund13}\\
1234 & $2455797.74568^{+0.00032}_{-0.00031}$ & 0.0007187 & \citet{Kund13}\\
1249 & $2455817.33688^{+0.00041}_{-0.00041}$ & -0.0008739 & \citet{Vanko13}\\
1398 & $2456011.95934^{+0.00073}_{-0.00073}$ & -0.0001536 & \citet{Turnr13}\\
1400 & $2456014.57219^{+0.00070}_{-0.00069}$ & 0.0003241 & \citet{Turnr13}\\
1400 & $2456014.57248^{+0.00065}_{-0.00065}$  & 0.0006141 & \citet{Turnr13}\\
1411 & $2456028.93996^{+0.00049}_{-0.00049}$ & 0.0000462 & \citet{Turnr13}\\
1448 & $2456077.27003^{+0.00035}_{-0.00037}$ & 0.0014278 & this work\\
1452 & $2456082.49260^{+0.0012}_{-0.0011}$ & -0.0009469  & \citet{Pusk17}\\
1455 & $2456086.41124^{+0.00095}_{-0.0010}$ & -0.0008654 & \citet{Pusk17}\\
1465 & $2456099.47337^{+0.0016}_{-0.0016}$ & -0.0005972 & this work\\
1690 & $2456393.36471^{+0.00048}_{-0.00050}$ & -0.0011460 & this work\\
1713 & $2456423.40717^{+0.00079}_{-0.00079}$ & -0.0009679 & this work\\
1719 & $2456431.24526^{+0.00032}_{-0.00031}$ & 0.0000050 & this work\\
1726 & $2456440.38874^{+0.00063}_{-0.00063}$ & 0.0001818 & \citet{Pusk17}\\
1762 & $2456487.41277^{+0.00075}_{-0.00075}$ & 0.0015096 & \citet{Pusk17}\\
1961 & $2456747.34245^{+0.00019}_{-0.00019}$ & 0.0001413 & this work\\
2033 & $2456841.38981^{+0.00078}_{-0.00079}$ & 0.0020969 & \citet{Pusk17}\\
2262 & $2457140.50425^{+0.00041}_{-0.00039}$ & -0.0000966 & \citet{Pusk17}\\
2311 & $2457204.50652^{+0.00087}_{-0.00089}$ & -0.0009490 & \citet{Pusk17}\\
2317 & $2457212.34484^{+0.00052}_{-0.00047}$ & 0.0002539 & \citet{Pusk17}\\
2324 & $2457221.48724^{+0.00062}_{-0.00063}$ & -0.0006493 & \citet{Ricc17}\\
2327 & $2457225.40653^{+0.00036}_{-0.00034}$ & 0.0000822 & \citet{Pusk17}\\
2337 & $2457238.46724^{+0.00064}_{-0.00066}$ & -0.0010695 & \citet{Ricc17}\\
2340 & $2457242.38679^{+0.00025}_{-0.00025}$ & -0.0000780 & \citet{Pusk17}\\
2351 & $2457256.75464^{+0.00072}_{-0.00074}$ & -0.0002759 & \citet{Ricc17}\\
2353 & $2457259.36698^{+0.00058}_{-0.00057}$ & -0.0003083 & \citet{Pusk17}\\
2531 & $2457491.86832^{+0.00040}_{-0.00041}$ & -0.0001070 & \citet{Ricc17}\\
2570 & $2457542.80916^{+0.00026}_{-0.00026}$ & -0.0005277 & \citet{Ricc17}\\
3062 & $2458185.45419^{+0.00060}_{-0.00064}$ & 0.0009055 & this work\\
3065 & $2458189.37246^{+0.00068}_{-0.00065}$ & 0.0006169 & this work\\
3075 & $2458202.43178^{+0.0015}_{-0.0015}$ & -0.0019248 & this work\\
3078 & $2458206.35282^{+0.00076}_{-0.00071}$ & 0.0005567 & this work\\
3088 & $2458219.41292^{+0.00057}_{-0.00058}$ & -0.0012050 & this work\\
3091 & $2458223.33346^{+0.00053}_{-0.00050}$ & 0.0007765 & this work\\
\hline
\end{longtable*}
{Note: $^a$ Mid-transit time ($T_m$) directly taken from \citet{Jiang13}.}
\end{center}

\begin{figure*}
		\includegraphics[width=\columnwidth]{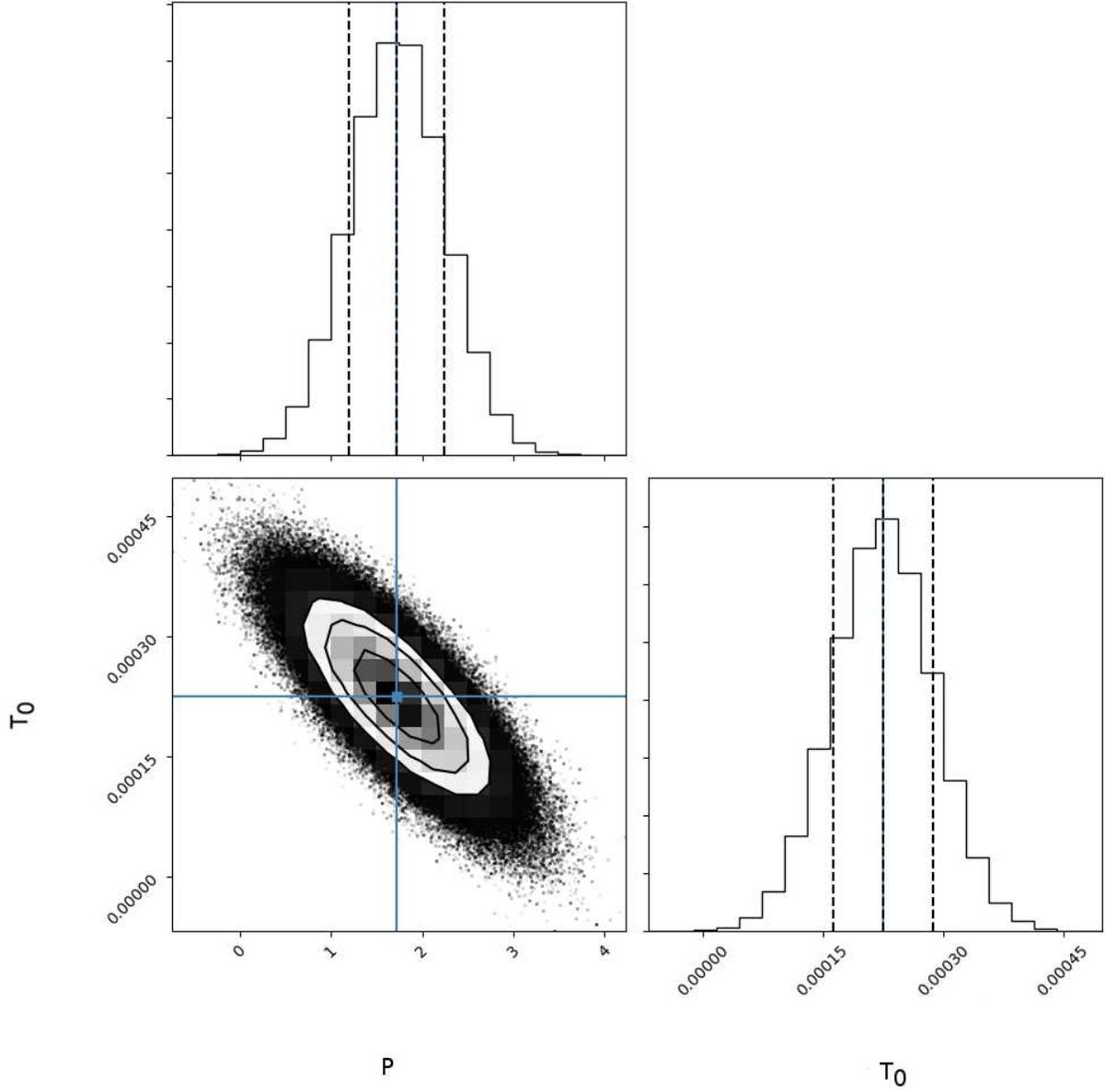}
    \caption{Corner plot depicting the marginalized 1-D and 2-D posterior probability distributions for the parameters of linear ephemeris model. The diagonal panel shows marginalized 1-D posterior probability distribution of each parameter, where the three vertical dashed lines from left to right represent the 16.0 (lower $1\sigma$ uncertainty), 50.0 (median as best fit value of model parameter) and 84.0 (upper $1\sigma$ uncertainty) percentile levels of the posterior probability distribution, respectively. The off-diagonal panel shows the marginalized 2-D posterior probability distribution for pair of parameters, with $1\sigma$, $2\sigma$, and 3$\sigma$ credible intervals marked with black contours. The gray shading corresponds to probability density (darker for higher probability). The solid vertical and horizontal blue lines show the best-fit values of the model parameters. The scaling of the parameters $P$ and ${T}_{0}$ on the above panels should be considered as $P= P \times {10}^{-7}+1.306186 \ (days)$ and ${T}_{0}={T}_{0}+2454185.911$ $({BJD}_{TDB})$, respectively.}
    \label{fig:2}
\end{figure*}

\subsection{A Search for Periodicity in the Timing Residuals through Frequency Analysis}
In order to search for periodicity in the timing residuals given in Table 6, we computed a generalized Lomb-Scargle periodogram \citep[GLS;][]{Zechme09} in the frequency domain. The periodogram defined by the resulting spectral power as function of frequency is shown in Figure 4. In this periodogram, we found the highest power peak (power=$0.1383$) at the frequency of $ 0.043867 \ cycle/period$. The False Alarm Probability (FAP) of ${26\%}$ for the highest power peak was determined empirically by randomly permuting the timing residuals to the observing epochs using a bootstrap resampling method with ${10}^5$ trials. As shown in Figure 4, this FAP of highest power peak is found to be far below the threshold levels of FAP=5\% and 1\%. This indicates that the presence of possible TTV in TrES-3 system does not show any signature of periodicity. Since there is no evidence of short-term TTV due to lack of periodicity in the timing residuals, it encouraged us to look for the long-term TTV that may be produced by either orbital decay or apsidal precession phenomenon in TrES-3 system. \\

\begin{figure*}[ht]
	\begin{center}
	\includegraphics[width=\columnwidth]{Fig_rev_3.eps}
     \caption{O-C diagram for analysis of all eighty three mid-transit times considered in this work. Black filled up-triangles are for the data from \citet{Sozz09}, red filled down-triangles are for \citet{Gibs09}, open magenta diamond is for \citet{Coln10}, turquoise filled left-triangles are for \citet{Lee11},  green filled squares are for \citet{Jiang13}, indigo filled squares are for \citet{Turnr13}, maroon filled right-triangles are for \citet{Kund13}, filled blue diamonds are for \citet{Vanko13},  open circles are for \citet{Pusk17}, violet filled star are for \citet{Ricc17} and black filled circle are for this work. The dashed red and blue curves indicate the timing residuals of orbital decay and apsidal precession ephemeris models, respectively.}
    \label{fig:3}
	\end{center}
\end{figure*}

\begin{figure*}[ht]
	\begin{center}
	\includegraphics[width=\columnwidth]{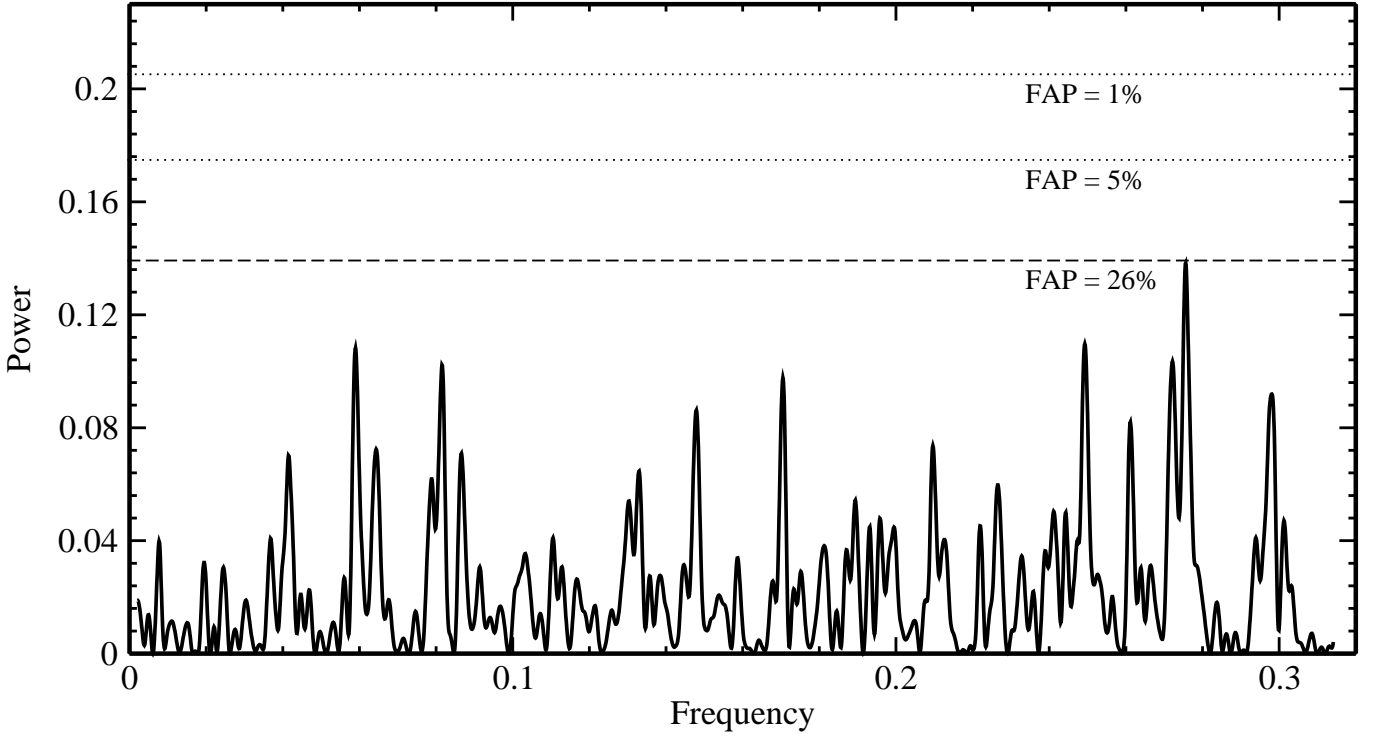}
     \caption{Generalized Lomab-Scargle Periodogram for eighty three timing residuals of TrES-3b. The dashed line indicates the FAP level of the highest peak of frequency 0.043867 cycle/period. The dotted lines from top to bottom indicate the threshold levels of FAP = 1\% and FAP = 5\%, respectively.}
    \label{fig:4}
	\end{center}
\end{figure*}

\subsection{Orbital Decay Study of TrES-3b}
Since \citet{Levrd09}, \citet{Matsu10}, and \citet{Penev18} have predicted TrES-3b to be tidally unstable and migrating inward towards its parent star due to tidal orbital decay, we made an attempt to explore this phenomenon in TrES-3 system using the transit time data spanning over a decade. For this study, we followed \citet{Adam10}; \citet{Blecc14}; and \citet{Jiang16} and constructed the orbital decay ephemeris model by adding a quadratic term to the linear ephemeris model. As similar to the linear ephemeris model fit (see Section 4.1), we used the $emcee$ MCMC sampler technique to fit the eighty three mid-transit times ${T}_{m}$ as a function of epoch E given in Table 6 to the following orbital decay ephemeris model:
\begin{equation}
T^{c}_{q}(E) = {T}_{q0} + P_qE + \delta P \frac{E(E - 1)}{2},           
\end{equation}
where E is the epoch, ${T}_{q0}$ is the mid-transit time at E=0, $P_q$ is the orbital period, $\delta P$ is the change of orbital period in each orbit, and $T^{c}_{q}(E)$ is the calculated mid-transit time. In order to estimate the best-fit values for the parameters $P_q$, ${T}_{q0}$ and $\delta P$ of the orbital decay ephemeris model, we considered uniform prior and Gaussian likelihood to sample their posterior probability distributions. The uniform prior used for each parameter is listed in Table 7. The number of walkers and the procedure of discarding some steps of every walker as the initial and final burn-in were exactly the same as those adopted in the linear ephemeris model fit (see Section 4.1). In order to have large effective number of independent samples, we ran 32,000 steps per walker of MCMC, which was larger than those considered in the linear ephemeris model fit. From this MCMC analysis, the estimated mean acceptance fraction, the integrated autocorrelation time and the effective number of independent samples were found to be $\sim 0.35$, $\sim 30$ and $\sim 1066$, respectively. The discussed acceptability and reliability of these estimated parameters in Section 4.1 indicate the well performance and convergence of MCMC chain. After discarding 60 steps (i.e., nearly two times the estimated value of integrated autocorrelation time) as a final burn-in from 32,000 steps per walker of MCMC, the remaining samples of model parameters $P_q$, ${T}_{q0}$ and $\delta P$ were used for Bayesian parameter extraction. The estimated best-fit values of these parameters along with their $1\sigma$ uncertainties (i.e., the medians and 68\% credible intervals of the posterior probability distributions) are given in Table 7. The corner plot depicting the marginalized 1-D and 2-D posterior probability distributions for the parameters of orbital decay ephemeris model is shown in Figure 5. As similar to Figure 2, the marginalized 1-D posterior probability distributions for the parameters of orbital decay ephemeris model are found to be symmetric and Gaussian (diagonal panel). In addition to this, the best-fit model parameters are also found to be lying within the smooth and smaller in size $1\sigma$ contour (off-diagonal panel). This indicates that the fitting of orbital decay ephemeris model to the transit time data is reliable. The minimum ${\chi}^{2}$ of this model fit is $148.24$, ${\chi}^{2}_{red}(80)$ is $1.853$, and the value of BIC is $161.41$. Using the best-fitted orbital decay ephemeris given in Table 7, the ${T}^{c}_{q} (E)$ was calculated for each epoch E. By subtracting the mid-transit times calculated using linear ephemeris, ${T}^{c}_{m}(E)$, from the above estimated ${T}^{c}_{q} (E)$, the timing residual ${T}^{c}_{q} (E)$-${T}^{c}_{m}(E)$ of orbital decay ephemeris model was obtained and plotted as a function of epoch E with red dashed curve in Figure 3. The $RMS$ of the timing residuals is found to be $\sim 70.03 \ s$. Using the values of parameters $P_q$ and $\delta P$ given in Table 7, the decay rate \citep[$\dot{P_q}=\frac{\delta P}{P_q}$,][]{Jiang16} of orbital period of TrES-3b is found to be $ \sim -4.1 \pm 3.1 \ ms\ {yr}^{-1}$.
\begin{figure*}
		\includegraphics[width=\columnwidth]{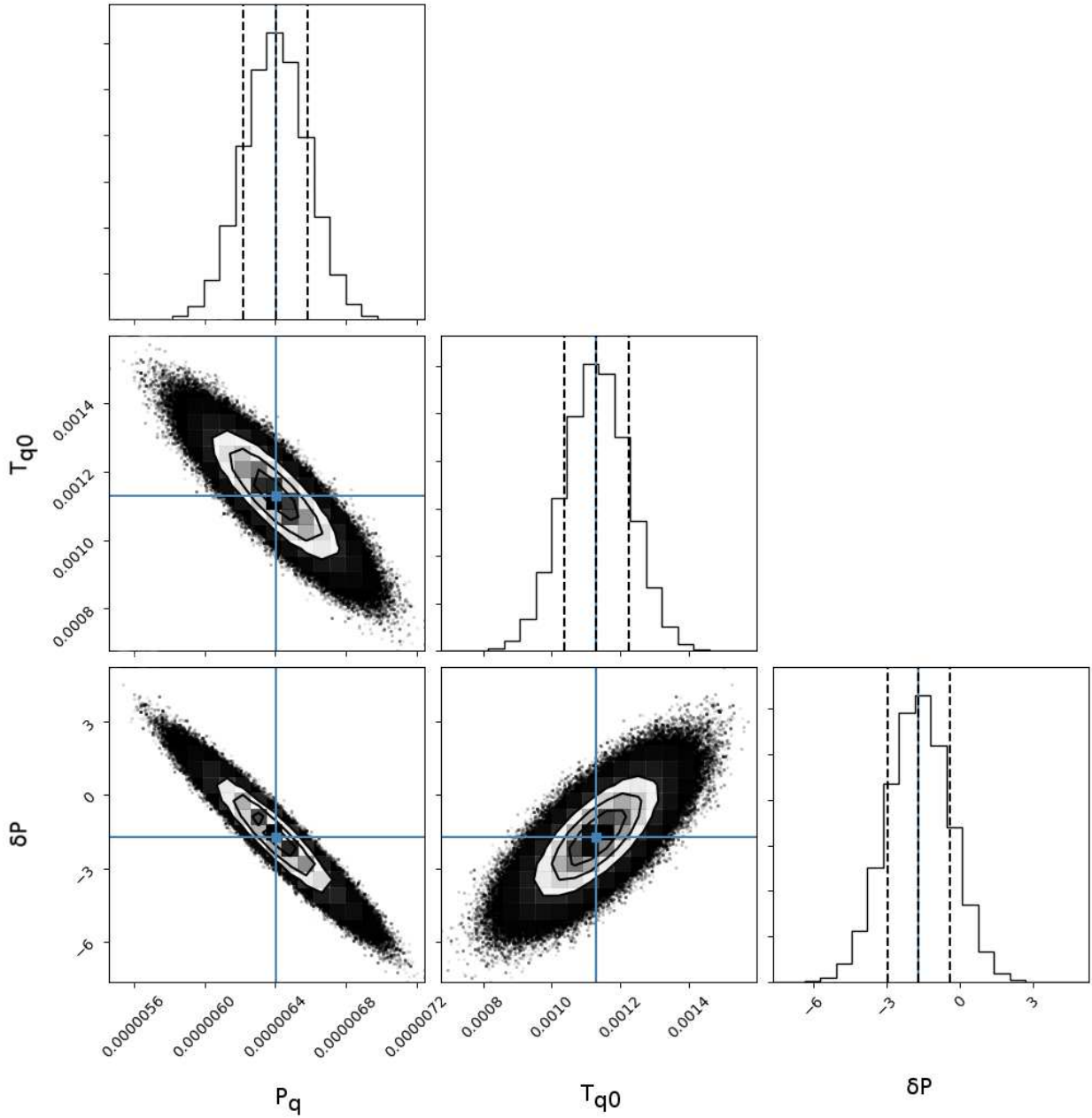}
    \caption{Corner plot depicting the marginalized 1-D and 2-D posterior probability distributions for the parameters of orbital decay ephemeris model. The other features of the panels are exactly the same as those mentioned in Figure 2. The scaling of the parameters $P_q$, ${T}_{q0}$ and $\delta P$ on the above panels should be considered as ${P}_{q}={P}_{q}+1.30618 \ (days)$, ${T}_{q0}={T}_{q0}+2454185.91$ $({BJD}_{TDB})$ and $\delta P={\delta P} \times {10}^{-10} \ (days)$, respectively} 
    \label{fig:5}
\end{figure*}

\subsection{Apsidal Precession Study in TrES-3 System}
\citet{Ragoz09} have already suggested TrES-3b to be a potential candidate to examine the apsidal precession as long as its orbit is at least slightly eccentric with $e>0.003$. Therefore, we probed the possibility of this phenomenon in TrES-3 system by adopting the following apsidal precession ephemeris model derived from Equations (7), (9), and (10) of \citet{Patra17}: 
\begin{equation}
T^{c}_{ap} (E) = T_{ap0} + P_{s} E -  \frac{e {P_s} \cos{({\omega}_0 + E \frac{d\omega}{dE})}}{\pi(1-\frac{\frac{d\omega}{dE}}{2 \pi})},
\end{equation}    
where E is the epoch, $T^{c}_{ap}(E)$ is the calculated mid-transit time, $P_{s}$ is the sidereal period, $e$ is the eccentricity of orbit, $\omega$ is the argument of periastron, $\omega_0$  is the argument of periastron at epoch zero (E=0), and $\frac{d\omega}{dE}$ is the precession rate of periastron. By following previous two model fits (see Section 4.1 and 4.3), we used $emcee$ MCMC sampler technique and fitted the above Equation (3) representing apsidal precession ephemeris model to the eighty three mid-transit times given in Table 6 as a function of epoch E. In this MCMC analysis, the uniform prior and Gaussian likelihood were assumed to sample the posterior probability distributions for the parameters ${P}_{s}$ ${T}_{ap0}$, $e$, $\omega_0$ and $\frac{d\omega}{dE}$ of apsidal precession ephemeris model. The uniform prior used for each of these model parameters is listed in Table 7. We followed linear ephemeris model fit and used the same number of walkers, as well as adopted exactly the same procedure of discarding some steps of every walker as the initial burn-in. However, we ran $2 \times {10}^{5}$ steps per walker of MCMC in order to have sufficient effective number of independent samples of model parameters. During the initial test runs of model fits, we noticed that when the uniform prior for eccentricity $e$ was assumed to be either (0, 0.05) or (0, 0.01), the $1 \sigma$ uncertainty in mid-transit time ${T}_{ap0}$ was found to be increased by 1 or 2 order more as compared to those obtained in the previous two model fits with the ${\chi}^{2}_{red}>6$. After several test runs of model fits, the slightly improved fit was achieved when the uniform prior for $e$ was assumed to be (0, 0.003). In this model fit, the estimated values of mean acceptance fraction, integrated autocorrelation time and the effective number of independent samples were found to be $\sim 0.23$, $\sim 782$ and $\sim 255$, respectively. The good sampling and convergence of MCMC chain are suggested by the acceptability and reliability of these estimated parameters discussed in Section 4.1. However, the rate of convergence appears to be slow due to the larger value of integrated autocorrelation time of $\sim 782$ steps as compared to previous two model fits (see Section 4.1 and 4.3). This suggests that the samples drawn before $ \sim 782$ steps have strongly correlated model parameters. In order to avoid this, the initial 2346 steps (i.e., nearly three times the estimated value of integrated autocorrelation time) were discarded as a final burn-in from the $2 \times {10}^{5}$ steps per walker of MCMC. Finally, Bayesian parameter extraction was performed using the remaining samples of model parameters ${P}_{s}$ ${T}_{ap0}$, $e$, $\omega_0$ and $\frac{d\omega}{dE}$. The best-fit values of these parameters along with their $1\sigma$ uncertainties (i.e., the medians and the 68\% credible intervals of the posterior probability distributions) are listed in Table 7. The corner plot depicting the marginalized 1-D and 2-D posterior probability distributions for the parameters of apsidal precession ephemeris model is shown in Figure 6. In contrast to Figure 2 and Figure 5, the marginalized 1-D posterior probability distribution for each parameter of apsidal precession ephemeris model does not appear to be symmetric and Gaussian (diagonal panel). Besides this, the marginalized 2-D posterior probability distribution for each pair of model parameters seems to be asymmetric and broader and for most of the cases, the best-fit values of model parameters lie outside the $1\sigma$ contour (off-diagonal panel). In this MCMC analysis, the measured values of the model parameters $e$, $\omega_0$ and $\frac{d\omega}{dE}$ appear to be statistically less significant as the estimated $1\sigma$ uncertainties in these parameters are found to be large (see Table 7). All these obtained unusual features might be originated due to nearly circular orbit of TrES-3b with ${e=0.00137^{+0.00105}_{-0.00092}}$ (see Table 7), as well as the strongly correlated model parameters. This also indicates that the fitting of apsidal precession ephemeris model to the transit time data is not much reliable. The minimum ${\chi}^{2}$ of this model fit is ${220.27}$, ${\chi}^{2}_{red} (78)$ is ${ 2.824}$, and the value of BIC is ${242.39}$. Using the best-fitted apsidal precession ephemeris given in Table 7, the mid-transit time ${T}^{c}_{ap} (E)$ was calculated for each epoch E. By subtracting the mid-transit times calculated using linear ephemeris, ${T}^{c}_{m}(E)$, from the above estimated ${T}^{c}_{ap} (E)$, the timing residual ${T}^{c}_{ap} (E)$-${T}^{c}_{m}(E)$ of apsidal precession ephemeris model was obtained and plotted as a function of epoch E with blue dashed curve in Figure 3. The $RMS$ of this timing residuals is found to be $\sim 69.91\ s$.

\begin{figure*}
		\includegraphics[width=\columnwidth]{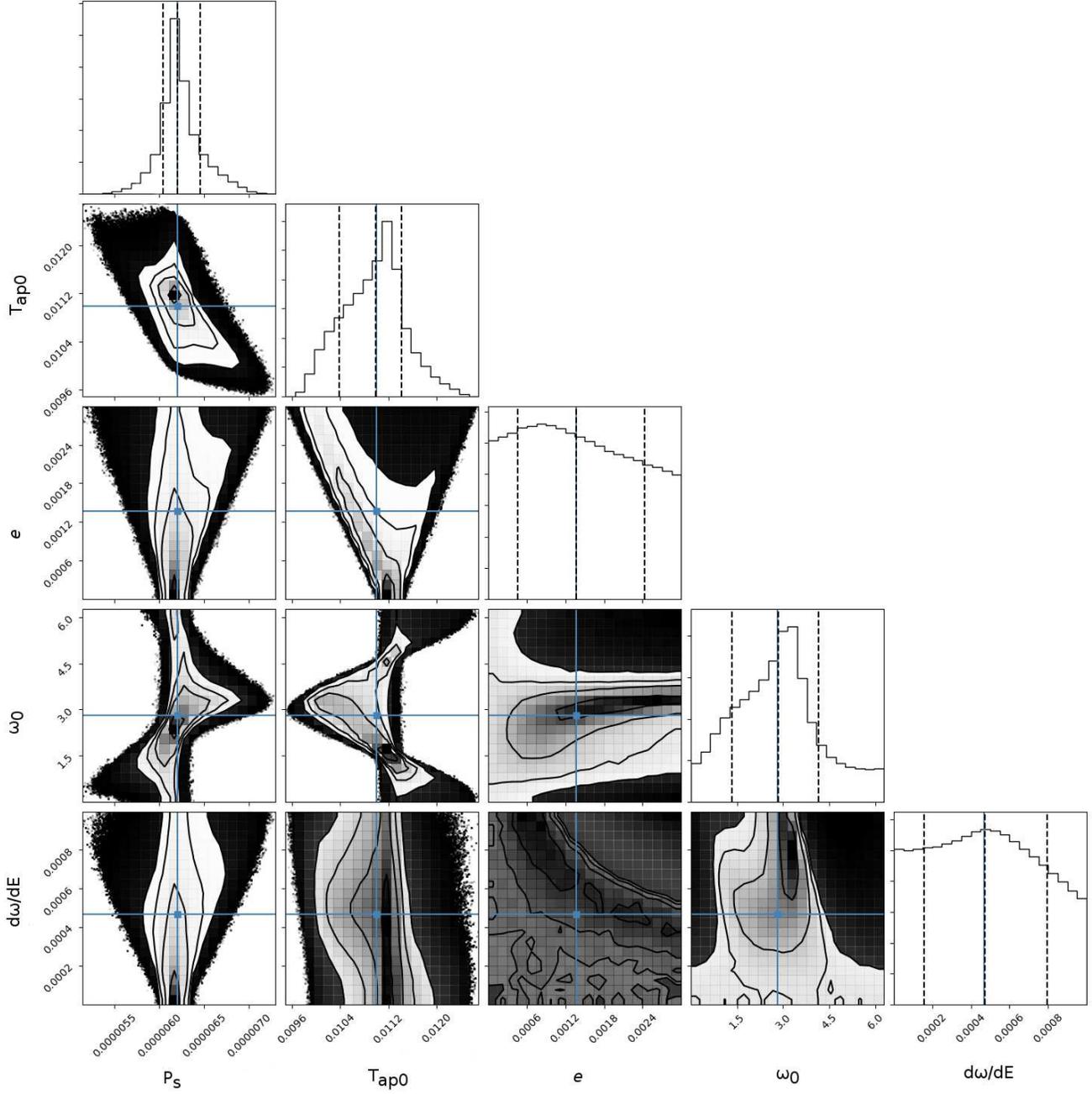}
    \caption{Corner plot depicting the marginalized 1-D and 2-D posterior probability distributions for the parameters of apsidal precession ephemeris model. The other features of the panels are exactly the same as those mentioned in Figure 2. The scaling of the parameters $P_s$ and ${T}_{ap0}$ on the above panels should be considered as ${P}_{s}={P}_{s}+1.30618 \ (days)$ and  ${T}_{ap0}={T}_{ap0}+2454185.9$ $({BJD}_{TDB})$, respectively.}   
    \label{fig:5}
\end{figure*}

\begin{table*}[ht]
\begin{center}
\caption {The Uniform Priors and Best-fit Model Parameters}
\label{tab:3}
\small\addtolength{\tabcolsep}{-2pt}
\begin{tabular}{llcc}
\hline
\hline
Model & Parameter & Uniform Prior & Best-fit Model Parameter\\
\hline
Linear ephemeris & $P$ [days] & (0, 2) & ${1.306186172}^{+0.000000052}_{-0.000000052}$\\
& ${T}_{0}$ [${BJD}_{TDB}$] & (2454184, 2454186) & ${2454185.911225}^{+0.000062}_{-0.000062}$\\
Orbital decay ephemeris & $P_q$ [days] & (0, 2) & ${1.306186401}^{+0.000000181}_{-0.000000180}$\\
& ${T}_{q0}$ [${BJD}_{TDB}$] & (2454184, 2454186) & ${2454185.911131}^{+0.000094}_{-0.000094}$\\
&  ${\delta P}$ [days]$^{a}$ & (-1, 1) & ${-1.702171}^{+1.283450}_{-1.286600}$ \\
Apsidal Precession ephemeris & ${P}_{s}$ [days]  & {(0, 2)} & ${1.306186199}^{+0.000000253}_{-0.000000160}$ \\
& ${T}_{ap0}$ [${BJD}_{TDB}$] & (2454184, 2454186) & ${2454185.910992}^{+0.000424}_{-0.000691}$\\
& $e$ & (0, 0.003) & ${0.00137}^{+0.00105}_{-0.00092}$\\
& ${\omega_0}$ [rad] & (0, 2$\pi$) & ${2.838}^{+1.306}_{-1.510}$\\
& $\frac{d\omega}{dE}$ [${rad}\ {epoch}^{-1}$] & (0, 0.001) & ${0.000472}^{+0.000323}_{-0.000314}$\\
\hline
\end{tabular} \\
{Note: $^{a}$ The uniform prior for $\delta P$ is in days, while its best-fit value is in ${10}^{-10}$ days}
\end{center}
\end{table*}

\section{Discussion: Implications of Ephemeris Models} 
\subsection{Most Plausible Model Representing the Transit Time Data}
As several authors \citep[e.g.,][]{Blecc14,Hoyer16a,Hoyer16b,Patra17} have adopted BIC statistic for obtaining the most plausible model that can represent the transit time data, we also adopted the same procedure and calculated the values of BIC corresponding to minimum values of ${\chi}^2$ obtained from linear, orbital decay, and apsidal precession ephemeris model fits. The obtained values of BIC from the orbital decay and apsidal precession ephemeris model fits (see Section 4.3 and 4.4) favor the orbital decay of TrES-3b rather than the apsidal precession by $\bigtriangleup BIC = {80.98}$, corresponding to the approximate Bayes factor of  $\exp(\bigtriangleup BIC/2) = {3.84 \times {10}^{17}}$ \citep[cf.][]{Blecc14,Maci16,Patra17}. This is also justified by the fact that the very low value of eccentricity $e=0.00137^{+0.00105}_{-0.00092}$, estimated from the apsidal precession ephemeris model fit (see Table 7), would have a marginal effect in the transit parameters determined from the transit light curves \citep{Maci16}. Moreover, the unusual features obtained while fitting the apsidal precession ephemeris model to the considered transit time data also favor to rule out possibility of this phenomena in TrES-3 system (see Section 4.4). In this regard, it is worth mentioning here that the observations of secondary eclipses would be important to provide tight constraint on the eccentricity and apsidal precession rate of TrES-3 system. Since the values of ${\chi}^{2}_{red}$ and BIC are not significantly different between the linear and orbital decay ephemeris model fits (see Section 4.1 and 4.3), it is difficult to find which one of these two models can represent the transit time data considered in this work. However, we prefer linear ephemeris model over the orbital decay ephemeris model for the presently available transit time data by considering the slightly smaller value of BIC for the former model fit in comparison to the later one, as well as the measured orbital decay rate is consistent within $\sim 1.3 \sigma$, indicating its statistically less significant estimation (see Section 4.3). In this context, it is noteworthy here that further follow-up observation of transits of TrES-3b may provide the statistically possible measurement of orbital decay rate and thus would be useful to rule in or out this phenomenon. 

\subsection{Estimation of ${Q}^{'}_{\ast}$ of TrES-3}
Assuming that the measured decreasing period of TrES-3b is real and attributed to orbital decay, the modified stellar tidal quality factor (${Q}^{'}_{\ast}$) that indicates the efficiency of tidal dissipation in the host star, is estimated using the following equation of \citet{Maci16}:
\begin{equation}
{Q}^{'}_{\ast} = 9{P_q}{{\dot{P_q}}^{-1}}\left(\frac{M_p}{M_\ast}\right)\left(\frac{a}{R_\ast}\right)^{-5}\left(\omega_\ast - \frac{2\pi}{P_q}\right),
\end{equation}
where $P_q$ is the orbital period, $\dot{P_q}$ is the decay rate of orbital period, $\frac{M_p}{M_\ast}$ is the mass ratio of planet to star, $\frac{a}{R_\ast}$ is the ratio of semi-major axis to stellar radius, and $\omega_\ast$ is the rotational frequency of the host star. The values of $P_q$ and $\dot{P_q}$ are already calculated in Section 4.3. However, the values of $\frac{M_p}{M_\ast}$ = 0.001964, and $\frac{a}{R_\ast} = 5.926$ are taken from \citet{Sozz09}, whereas $\omega_\ast = 0.2294808 \ {day}^{-1}$ is calculated from the stellar rotation period given in \citet{Matsu10}. We substituted these values in the above Equation (4) and found the modified stellar tidal quality factor to be ${Q}^{'}_{\ast} \sim 1.11$ $\times {10}^5$ for TrES-3, which lies within the typical range of ${10}^{5}-{10}^{7}$ reported for the stars hosting the hot-Jupiters \citep{Essick16,Penev18}, and is also of the same order of magnitude as found by \citet{Sun18}. As the timescale of orbital evolution of hot-Jupiters depends on the efficiency of tidal dissipation in their host stars, we substituted ${Q}^{'}_{\ast} \sim 1.11 \times {10}^{5}$ and relevant parameters from \citet{Sozz09} in the following equation of \cite{Levrd09} to calculate the remaining lifetime of TrES-3b (before it collides with its host star):
\begin{equation}
T_{remain} = {\frac{1}{48}\frac{{Q}^{'}_{\ast}}{n}\left(\frac{M_\ast}{M_p}\right)\left(\frac{a}{R_\ast}\right)^5},
\end{equation}
where $n=\frac{2 \pi}{P_q}$ is the frequency of mean orbital motion of the planet and found its value to be ${T}_{remain} \sim 4.9 \ Myr$. In order to estimate the expected shift in the transit arrival time of TrES-3b due to its decaying orbit, we used the following equation of \citet{Birkby14}:
\begin{equation}
T_{shift} = {\frac{1}{2}}{T^2}\left(\frac{dn}{dT}\right)\left(\frac{P_q}{2\pi}\right) , 
\end{equation}
where $\frac{dn}{dT}$ is the current rate of change of frequency of mean orbital motion of the planet whose expression can be obtained in terms of ${Q}^{'}_{\ast}$ by substituting $\dot{P_q} = \frac{P^{2}_{q}}{2\pi}\left(\frac{dn}{dT}\right)$ in the above Equation (4). For the calculated value of $\frac{dn}{dT} \sim 6.429 \times {10}^{-20}$ $rad \ s^{-2}$ corresponding to ${Q}^{'}_{\ast} \sim 1.11 \times {10}^{5}$, the expected shift in the transit arrival time of TrES-3b after eleven years (T=11 yr) is found to be ${T}_{shift} \sim 69.55 \ s$. This value of ${T}_{shift}$ is fully consistent with the $RMS$ of the obtained timing residuals shown in Figure 3. If ${Q}^{'}_{\ast} \sim 1.11 \times {10}^{5}$ measured from our timing analysis is maintained for another five years ({\it i.e.}, in a total of sixteen years monitoring), one can expect ${T}_{shift}$ $\sim 147 \ s$ that can be confirmed from the further follow-up observations. However, if ${Q}^{'}_{\ast} \sim {10}^{6}$, the expected ${T}_{shift}$ is only $\sim 16 \  s$ that appears to be difficult to detect.

\subsection{Estimations of Planetary Love Number ($k_{p}$)}
\citet{Ragoz09} showed for TrES-3 system that the contribution from planet's tidal deformation to the theoretical apsidal precession rate is larger as compared to that from the star's tidal deformation and general relativity. The rate of this precession is proportional to the second order planetary Love number ($k_p$), a dimensionless parameter that gives information about interior density profile of the planets. In order to calculate $k_p$ for TrES-3b, we assumed that the observed precession rate is real and adopted the following equation of \citet{Patra17}:
\begin{equation}
\frac{d\omega}{dE}={15}{\pi}{k_p}\left(\frac{M_\ast}{M_p}\right)\left(\frac{R_p}{a}\right)^5,
\end{equation}
Using the value of $\frac{d\omega}{dE}$ calculated from timing analysis in Section 4.4 and the other relevant parameters from \citet{Sozz09}, we found the value of ${k}_{p}$ to be ${1.15 \pm 0.32}$. Although the estimated value of $k_p$ is larger than that of Jupiter \citep[$k_p=0.59$:][]{Wahl16}, its measurement appears to be statistically less significant due to the larger value of uncertainty. As the model parameters $e$, ${T}_{ap0}$, $\omega_0$ and $\frac{d\omega}{dE}$ have strongly correlated errors, this may be the reason behind the larger value of uncertainty in the estimation of $k_p$ \citep[see][]{Bouma19}. In order to provide more tight constraint on the estimation of ${k}_{p}$, future secondary eclipse observations would be required.

\section{Concluding Remarks}
We present twelve new transit light curves of TrES-3b observed from 2012 May to 2018 April using three telescopes. For the precise timing analysis, we combine these transit light curves with seventy one transit data available in the literature and analyze them uniformly. All the orbital parameters determined from our twelve new transit light curves are consistent with previous reported results. Using the mid-transit times determined from the total of eighty three transit light curves, we derive new ephemeris for the orbital period and mid-transit time of TrES-3b that are found to be in good agreement with the previous results available in the literature. The transit timing analysis indicates the possibility of TTV in this planetary system. However, there is no evidence of additional body due to lack of periodic TTV as obtained from the frequency analysis. We have also explored the possibility of long-term TTV that may be induced due to orbital decay and apsidal precession. From the orbital decay study of TrES-3b, we find the orbital decay rate of $\dot{P_q}$ $={-4.1 \pm 3.1} \ {ms} \ {yr}^{-1}$. Considering the tidal dissipation within the host star as the cause of this orbital decay rate, we derive the modified stellar tidal quality factor of  ${Q}^{'}_{\ast} \sim 1.11 \times {10}^5$ for TrES-3, which lies within the typical range of ${10}^{5}-{10}^7$ reported for the stars hosting the hot-Jupiters. By assuming ${Q}^{'}_{\ast} \sim {1.11 \times} {10}^{5}$, the expected ${T}_{shift}$ in the transit arrival time of TrES-3b after eleven years is found to be $\sim 69.55 \ s$, which is fully consistent with the obtained $RMS$ of the timing residuals. Besides this, the apsidal precession study of TrES-3 system gives the apsidal precession rate of $\frac{d\omega}{dE}={0.000472}^{+0.000323}_{-0.000314}$ rad ${epoch}^{-1}$. For this precession rate, the estimated value of ${k}_{P} = 1.15 \pm 0.32$ appears to be statistically less significant due to the larger value of uncertainty. For our considered transit time data, we do not find the possibility of apsidal precession in TrES-3 system due to the very low value of eccentricity $e={{0.00137}^{+0.00105}_{- 0.00092}}$, as well as the larger value of BIC obtained for the apsidal precession ephemeris model fit in comparison to the linear and orbital decay ephemeris model fits. In order to rule in or out this phenomenon in TrES-3 system, the observation of secondary eclipse would be required. Because of the slightly smaller value of BIC for the linear ephemeris model fit as compared to the orbital decay ephemeris model fit, as well as the statistically less significant estimation of orbital decay rate, we prefer the linear ephemeris model for the presently employed eighty three transit time data. However, the possibility of slow orbital decay in TrES-3 system cannot be completely ruled out due to the following reasons: (i) the ${T}_{shift}$ is consistent with the $RMS$ of the timing residuals and (ii) the values of ${\chi}^{2}_{red}$ and BIC are not significantly different between the linear and orbital decay ephemeris models. In order to confirm this, further high-precision and high-cadence follow-up observation of transits of TrES-3b would be required. In this regard, it is worth mentioning here that the expected Transiting Exoplanet Survey Satellite (TESS) observations of TrES-3b from  2020 May to 2020 July would be useful to improve our understanding of the orbit of this extra-solar planet.

\section*{Acknowledgments}

We thank the anonymous referee for useful comments that improve the quality of the paper. We also thank J. Z. Gazak, D. Foreman-Mackey, and D. Ragozzine for their valuable suggestions and discussions, which have been very much helpful in improving the paper. We thank the staff at IAO, Hanle and CREST (IIA), Hosakote, as well as at DFOT (ARIES), Nainital for providing support during the observations. The time allocation committees of the HCT, DFOT, and the AZT-11 are gratefully acknowledged for providing the observation times. PT and VKM acknowledge the University Grants Commission (UGC), New Delhi for providing the financial support through Major Research Project no. UGC-MRP 43-521/2014(SR). PT expresses his sincere thanks to IUCAA, Pune for providing the supports through IUCAA Associateship Programme. IGJ acknowledges funding from the Ministry of Science and Technology, Taiwan, through the grand No. MOST 106-2112-M-007-006-MY3. YCJ acknowledges the Department of Science and Technology (DST), India for their  support through the Indo-Austria project on transiting exoplanets "INT/AUSTRIA/BMWF/P-14". MV would like to thank the project VEGA 2/0031/18 and APVV-15-0458. \c{C}P acknowledges the funding from TUBITAK (Scientific and Technological Research Council of Turkey) under Grant No. 113F353. \c{C}P also thanks Canakkale Onsekiz Mart University Astrophysics Research Center, Ulupinar Observatory, and Istanbul University Observatory Research and Application Center for their support and allowing use of T122 and T60 which were supported partly by National Planning Agency (DPT) of Turkey (project DPT-2007K120660 carried out Canakkale Onsekiz Mart University) and the Scientific Research Projects Coordination Unit of Istanbul University (project no. 3685). \c{C}P thanks to TUBITAK for a partial support in using T100 telescope with project numbers 13CT100-523 and 13CT100-537. We thank N. P. Gibson, J. W. Lee, and D. Ricci for sharing the transit light curves of TrES-3b with us. We also thankful to A. Sozzetti, K. D. Col\'{o}n, P. Kundurthy, and J. D. Turner for making their transit light curves publicly available. 

\software{IRAF \citep{Tody86,Tody93}, emcee \citep{Foreman13}, corner \citep{Foreman16}, astroML \citep{VanderPlas12}, NumPy \citep{vander11}, SciPy \citep{Jones01}, hjd2bjd \citep{East10}, JKLTD \citep{South15}, TAP \citep{Gaz12}}

\end{document}